\documentclass[12pt]{article}
\usepackage{amssymb}
\usepackage{amsmath}
\usepackage{graphicx,color}

\usepackage{float}
\usepackage{graphics}
\usepackage{color}
\usepackage{amssymb,amsmath}
\usepackage{slashed}
\usepackage{epstopdf}
\usepackage{booktabs}
\usepackage{mathrsfs}
\usepackage[export]{adjustbox}
\usepackage{epsfig}
\usepackage{graphicx,color}

\newcommand{\bea}{\begin{eqnarray}}
\newcommand{\eea}{\end{eqnarray}}
\newcommand{\sy}{s(y)}

\numberwithin{equation}{section}

\begin{document}
%
\begin{titlepage}
\vspace*{10mm}
\begin{center}
\baselineskip 25pt 
{\Large\bf
Fermion Mass Hierarchy and Phenomenology in the 5D Domain Wall Standard Model  

}
\end{center}
\vspace{5mm}
\begin{center}
{\large
Nobuchika Okada$^1$, 
Digesh Raut$^2$, and  
Desmond Villalba$^3$
}

\vspace{.5cm}

{\baselineskip 20pt \it
$^1$Department of Physics and Astronomy, \\
The University of Alabama, Tuscaloosa, AL 35487, USA\\
} 
{\baselineskip 20pt \it
$^2$Department of Physics and Astronomy, \\
The University of Delaware, Newark , DE 19716, USA\\
} 
{\baselineskip 20pt \it
$^3$Department of Chemistry and Physics, \\
Drury University, Springfield , Mo 65802, USA\\
}
\end{center}
\vspace{0.5cm}
\begin{abstract}

We have recently proposed a setup of the ``Domain-Wall Standard Model'' in 5D spacetime, 
   where all the Standard Model (SM) fields are localized in certain domains of the extra 5th dimension. 
Utilizing this setup, we attempt to solve the fermion mass hierarchy problem of the SM. 
The mass hierarchy can be naturally explained by suitably distributing the fermions in different positions along the extra dimension. 
Due to these different localization points, the effective 4D gauge couplings of Kaluza-Klein (KK) mode gauge bosons 
   to the SM fermions become non-universal. 
As a result, our model is severely constrained by the Flavor Changing Neutral Current (FCNC) measurements. 
We find two interesting cases in which our model is phenomenologically viable: 
   (1) the KK-mode of the SM gauge bosons are extremely heavy and unlikely to be produced at the Large Hadron Collider (LHC), 
      while future FCNC measurements can reveal the existence of these heavy modes.~(2) the width of the localized SM fermions is very narrow, leading to almost universal 4D KK-mode gauge couplings.  
   In this case, the FCNC constraints can be easily avoided even if a KK gauge boson mass lies at the TeV scale. 
   Such a light KK gauge boson can be searched at the LHC in the near future.

\end{abstract}
\end{titlepage}

\section{Introduction}

The notion that our Universe may consist of more than 3-spatial  dimensions has enamored us for some time. 
The discovery of the D-branes in string theory \cite{D-brane} has provided theorists with a rich theoretical framework to construct theories with extra dimensions, 
    which has resulted in many applications towards resolving the shortcomings of the Standard Model (SM).  
The large extra-dimension model \cite{ADD} is a well-known brane-world scenario which offers a solution 
    to the gauge hierarchy problem, where the original Planck mass at the TeV scale 
    reproduces the 4D Planck mass by a large extra-dimensional volume. 
 Another well-known scenario is the warped extra-dimension model \cite{RS},  
    where the Planck scale is ``warped down'' to the TeV scale due to the anti-de Sitter (AdS) curvature 
    in the extra 5th dimension.

These well-known extra-dimensional models usually consist of compactified extra dimensions 
    on manifolds or orbifolds, and thereby treat the extra dimensions differently from the 3-spatial ones we reside in. 
We may think that a democratic choice regarding the extra dimension is more natural, 
    namely, all dimensions have the same feature of being flat and non-compact. 
This picture requires that all SM fields as well as 4D graviton are localized in certain domains in the bulk space. 
Historically the localization of scalar and fermion fields have been investigated in Ref.~\cite{DWF}. 
The localization of 4D graviton has been proposed in Ref.~\cite{RS2}, the so-called RS-2 scenario, 
    in the presence of the 5D AdS curvature. 
However, the localization of gauge fields has been known to be notoriously difficult. 
One successful attempt occurred about 20 years ago in Ref.~\cite{DS}. 
They considered a strongly coupled non-Abelian gauge symmetry in 5D bulk, 
    which is broken down to a smaller gauge group in domains where background defects persist. 
A very simple way for localizing the gauge field, which can be considered as an effective description of 
    the dynamical localization mechanism of the gauge field in Ref.~\cite{DS}, 
    has been proposed in Ref.~\cite{OS}, 
    where a gauge coupling depending on the extra dimensional coordinates plays the essential role for localizing the gauge field. 
Recently, the idea of Ref.~\cite{OS} has been refined in a theoretically consistent manner 
    with an extra-dimensional analog of the $R_\xi$ gauge \cite{DWSM, DWSM2, ABES, ABES2},  
    and the ``Domain-Wall Standard Model'' has been proposed, 
    where all the SM fields are localized in 
certain domains of the bulk space.

In this paper, one of the main pursuits is to explain the mass hierarchy among the SM fermions 
   by utilizing the setup of the 5D Domain-Wall SM in Ref.~\cite{DWSM2}. 
As has been originally proposed in Ref.~\cite{AS}, the fermion mass hierarchy 
    can be naturally explained by ``geometry'', namely, localizing the fermions
    in different positions along the extra dimension. 
In this ``split fermion scenario'', 
   the effective 4D Kaluza-Klein (KK) mode gauge couplings
    to the SM fermions become non-universal due to the different fermion localization points.  
As a result, the scenario is severely constrained 
    by the Flavor Changing Neutral Current (FCNC) measurements. 
Detailed analysis for the FCNC constraints has been performed in Ref.~\cite{LH}. 
Following their analysis, we consider the current FCNC constraints to identify the allowed parameter region of our model. 
We also consider the Large Hadron Collider (LHC) Run-2 results from the search for a narrow resonance 
    and interpret the LHC results into the constraints on the KK-mode gauge bosons. 
We find two separate scenarios to satisfy the experimental constraints: 
    (1) the KK-modes of the SM gauge bosons are extremely heavy and unlikely to be produced at the LHC, 
     while improved FCNC measurements in the near future can reveal the existence of these heavy modes. 
    (2) the width of the localized SM fermions is very narrow and the 4D KK-mode gauge couplings become almost universal. 
In this case, the FCNC constraints can be easily avoided even if a KK gauge boson mass is at the TeV scale.  
Such a light KK gauge boson can be searched at the LHC Run-3 and the High-Luminosity LHC in the near future.

This paper is organized as follows: 
In Sec.~\ref{sec:2} we review the gauge field localization procedure for the reader's convenience. 
In Sec.~\ref{sec:3} we summarize the localization procedure for the Higgs field and its vacuum expectation value (VEV) 
   which is exactly analogous to the gauge field localization mechanism. 
Section \ref{sec:4} focuses on the Domain-Wall (DW) SM fermion construction, 
   and we show how to reproduce the fermion mass hierarchy through ``geometry'' 
   without large hierarchies among the original model parameters. 
In Sec.~\ref{sec:5} we evaluate the 4D effective gauge couplings between the KK gauge bosons and the SM fermions. 
This aspect of our model is a crucial element for the phenomenology considerations in the next section, 
   as the non-universal nature of the gauge couplings induces FCNC processes. 
In Sec.~\ref{sec:6} we discuss how our model avoids the constraints from the FCNC measurements 
   induced by the non-universal gauge couplings. 
We also consider the LHC Run-2 results of the search for a narrow resonance 
   and interpret the results into the constraint on the KK-mode SM gauge bosons.    
We find that two distinct situations arise: either the KK gauge boson is very heavy, or it can be light enough to be produced at the LHC. 
We summarize our discussion in the last section.

\section{Domain-Wall Gauge Boson}
\label{sec:2}
Let us first consider the gauge sector of the Domain-Wall SM. 
Here, we review the localization procedure of the 5D gauge boson 
  which the the authors previously   introduced in Refs.~\cite{DWSM, DWSM2}.
The reader who is familiar with this procedure may skip this section. 
For simplicity, we consider a U(1) gauge theory.  
The Non-Abelian extension is completely analogous \cite{ABES}. 

We consider the following Lagrangian for the U(1) gauge field in 5D flat Minkowski space: 
\bea
\mathcal{L}_{5}=-\frac{1}{4}s(y)F_{MN}F^{MN}, 
\eea
where $F_{MN}$ is the gauge field strength, $M, N=0,1,2,3,y$ with $y$ being the index for the 5th coordinate, 
   and $g_{MN}={\rm diag}(1,-1,-1,-1,-1)$ is our convention for the metric. 
In the original Lagrangian we identify $s(y)=1/{\bar g}^2$ with $\bar{g}$ being a $y$-dependent 5D gauge coupling.\footnote{The authors have previously considered the origin of s(y) in \cite{DWSM2}. For the specific choice of s(y) given by Eq.~(\ref{exp2}), this 5D y-dependent gauge coupling can be expressed in terms of the kink solution (\ref{kink}) when $m_V=m_\phi$.} 
This $y$-dependence is the key for localizing the gauge field through a strong coupling 
   as has been first proposed in Ref.~\cite{OS}. 
In the 5D Lagrangian, the gauge field and $s(y)$ have a mass dimension of one.

Decomposing the field strength into its components yields the following expression (up to total derivative terms):
\bea
\mathcal{L}_{5}
= \frac{1}{2} s A^{\mu}\left(g_{\mu \nu}\Box_{4} -\partial_{\mu}\partial_{\nu}\right)A^{\nu}-\frac{1}{2} s A_{y}\Box_{4}A_{y} 
- \frac{1}{2}A_{\mu}\partial_{y}\left(s \partial_{y}A^{\mu}\right)-\left(\partial_{\mu}A^{\mu}\right)\partial_{y}\left(s A_{y}\right),
\label{L5} 
\eea
where $A_{\mu}$ $(\mu=0,1,2,3)$ and $A_y$ are a gauge field and a scalar field 
  in 4D space-time, and the first two terms in the right-hand side denote the kinetic terms for these fields.
The last term contains a mixing between $A_{\mu}$ and $A_y$, 
  which is analogous to the mixing term between a gauge field and a would-be Nambu-Goldstone (NG) mode 
  in spontaneously broken gauge theory if we regard $s A_{y}$ as the NG mode. 
Based on this observation, we introduce a gauge fixing term, which is a 5D analog to the $R_{\xi}$ gauge \cite{DWSM, ABES}: 
\bea
\mathcal{L}_{\rm GF}=-\frac{s}{2 \, \xi}\left(\partial_{\mu}A^{\mu}- \frac{\xi}{s} \partial_{y}(s A_y)\right)^2 ,
\eea
where $\xi$ is a gauge parameter. 
By this gauge fixing term, the mixing term between $A_{\mu}$ and $A_y$ is eliminated. 
The total Lagrangian now reads $\mathcal{L}=\mathcal{L}_{5}+\mathcal{L}_{\rm GF}=\mathcal{L}_{gauge}+\mathcal{L}_{scalar}$, 
where
\bea 
\mathcal{L}_{gauge}&=&\frac{1}{2} s A^{\mu}\left(g_{\mu \nu}\Box_{4}-\left(1-\frac{1}{\xi}\right)\partial_{\mu}\partial_{\nu}\right)A^{\nu}-\frac{1}{2}A_{\mu}\partial_{y}(s \, \partial_{y}A^{\mu}), 
\label{gauge}
\\ 
\mathcal{L}_{scalar}&=&-\frac{1}{2} s  A_{y}\Box_{4} A_{y}+\frac{1}{2} s \, \xi A_{y}\partial_{y} \left(\frac{1}{s}\partial_{y}(s A_{y}) \right).
\label{scalar}
\eea

Next, we analyze the KK-modes of the gauge and scalar fields via their mode expansions.
Using the KK-mode expansion for the gauge and scalar fields, 
\bea
A_{\mu}(x,y)=\sum _{n=0}^\infty A_{\mu}^{(n)}(x)\chi^{(n)}(y), \quad A_{y}(x,y)=\sum _{n=0}^\infty \eta ^{(n)}(x)\psi^{(n)}(y), 
\label{KK_EXP} 
\eea
we obtain the KK-mode equations from Eqs.~(\ref{gauge}) and (\ref{scalar}):  
\bea
\frac{d}{dy}
\left(s \,  \frac{d}{dy}\chi^{(n)} \right)+ s \, m_{n}^{2}\chi^{(n)}=0,  \; \; \; 
\frac{d}{dy}\left( \frac{1}{s} \frac{d}{dy} (s \psi^{(n)} ) \right)+ \tilde{m}_{n}^{2}\psi^{(n)}=0. 
\label{KK_EOM}
\eea
With the solutions of these KK-mode equations, the Lagrangians in Eqs.~(\ref{gauge}) and (\ref{scalar}) are written as 
\bea 
\mathcal{L}_{gauge}&=& \sum_{n=0}^\infty \frac{1}{2} s  \left( \chi^{(n)} \right)^2
 \left[ A_{\mu}^{(n)} \left(g^{\mu \nu}(\Box_{4} + m_n^2) -\left(1-\frac{1}{\xi}\right)\partial^{\mu}\partial^{\nu}\right)A_{\nu}^{(n)}  \right] , 
 \nonumber \\ 
\mathcal{L}_{scalar}&=&- \sum_{n=0}^\infty  \frac{1}{2} s   \left( \psi^{(n)} \right)^2  
  \left[ \eta^{(n)} \left( \Box_{4} + \xi \, \tilde{m}_n^2 \right) \eta^{(n)} \right].
\label{L2} 
\eea
For our analysis in this paper, we adopt a simple example for $s(y)$ considered in Ref.~\cite{DWSM2}: 
\bea 
  s(y)= f(y)^2 = \frac{M}{\left[ \cosh(m_V y) \right]^{2 \gamma}}, 
\label{exp2}  
\eea
where $M$ and $m_V$ are positive mass parameters, and $\gamma$ is a positive constant.

In solving the KK-mode equations, let us introduce the following new variables, 
  $\tilde{\chi}^{(n)}(y)$ and  $\tilde{\psi}^{(n)}(y)$, defined as  
\bea
  \tilde{\chi}^{(n)}(y) = f(y) \, \chi^{(n)}(y), \; \; \; \;  \tilde{\psi}^{(n)}(y) = f(y) \, \psi^{(n)}(y).
\label{newKK}  
\eea  
We then have the KK-mode equations of the form: 
\bea
&& \left[ -\partial_y^2 - \frac{\gamma (\gamma+1) m_V^2}{\cosh^2(m_Vy)} \right] \tilde{\chi}^{(n)} 
  = \left(m_n^2 - \gamma^2 m_V^2 \right)  \tilde{\chi}^{(n)}, \nonumber\\
&& \left[ -\partial_y^2 - \frac{\gamma (\gamma-1) m_V^2}{\cosh^2(m_Vy)} \right] \tilde{\psi}^{(n)} 
  = \left(m_n^2 - \gamma^2 m_V^2 \right)  \tilde{\psi}^{(n)}.  
\label{KK_EOM4}
\eea
These equations have the form of the 1D Schr\"odinger equation, 
  $(-\partial_y^2 +V) \Psi^{(n)}=E_n \Psi^{(n)}$. 
Since the potential corresponds to $V \propto -1/\cosh^2(m_V y) < 0$, 
  a bound state with $E_n <0$ exists. 
Although continuous modes exist for $E_n >0$, 
  they are not important in the following discussion and we do not consider them in this paper.  

We are interested in the localization of the gauge field, namely, bound states 
  from the  the Schr\"odinger equation satisfying 
  the following boundary conditions: $ |\tilde{\chi}^{(n)}(y)|< \infty$ for $y \to 0$,  
  and $\tilde{\chi}^{(n)}(y) \to 0$ for $|y| \to \infty$. 
Such solutions are described by using the hyper-geometric function $F[a,b; c; y]$ \cite{HO}. 
We find the eigenvalues for $\tilde{\chi}^{(n)}$ to be
\bea 
  m_n^2 = n \left( 2 \gamma -n \right) m_V^2  \; \; \; \; (n=0, 1, 2, \cdots < \gamma).
\label{mn1}  
\eea 
The number of (localized) KK-modes is terminated by a condition $E_n = m_n^2 - \gamma^2 m_V^2 < 0$, 
  and thus  a bound state is guaranteed to exist for $\gamma > 0$. 
The eigenfunctions for even numbers of 
  $n= 2 n^\prime \; (n^\prime =0,1,2, \ldots )$ and odd numbers of 
  $n= 2 n^{\prime \prime}+1 \; (n^{\prime \prime} =0,1,2,\ldots )$ 
are given by (up to normalization factor) 
\begin{eqnarray} 
 \tilde{\chi}^{(n^\prime)}(y) = \left[ \cosh(m_V y) \right]^{- \gamma} \, 
   F\left[-n^\prime , -\gamma + n^\prime; 1/2 ; 1- \cosh^2(m_V y) \right], 
\end{eqnarray}
and 
\begin{eqnarray}
  \tilde{\chi}^{(n^{\prime \prime})}(y) =  \sinh(m_V y) 
      \left[\cosh(m_V y) \right]^{- \gamma}  
  F\left[-n^{\prime \prime}, - \gamma + n^{\prime \prime}+1 ; 3/2 ; 1- \cosh^2(m_V y) \right],   
  \label{sol1} 
\end{eqnarray}
respectively. 

From Eq.~(\ref{KK_EOM4}), $\tilde{\psi}^{(n)}$ can be obtained by substituting $\gamma= \overline{\gamma}+1$ and $m_n^2 =\overline{m_n}^2 + (2 \overline{\gamma}+1)m_V^2 $ 
  into the first equation in Eq.~(\ref{KK_EOM4}). Thus the eigenvalues for $\tilde{\psi}^{(n)}$ are given by 
\bea 
  m_n^2 = (n+1) \left( 2 \gamma -(n+1) \right) m_V^2  \; \; \; \; (n=0, 1, 2, \cdots < \gamma-1).
\eea 
Note that no zero-mode exists for the scalar component $A_{y}$, 
  whereas we can see the pairing of the KK-mode mass spectrum 
  between the gauge fields and the corresponding would-be NG modes~\cite{DWSM}. 
For our discussion throughout of this paper, let us fix $\gamma=3$. 
In this case, we have only two KK-modes present with the mass eigenvalues $m_1^2=5 m_V^2$ and $m_2^2=8 m_V^2$.   
The canonically normalized KK-mode expansions are found to be
\bea
  &&A_{\mu}(x,y) =g \, A_{\mu}^{(0)}(x) + 2 \, g \, \sinh(m_V y)  \, A_{\mu}^{(1)}(x)+\, \frac{g}{\sqrt{5}} \, (5-4\cosh ^2(m_V y))  \, A_{\mu}^{(2)}(x), \nonumber \\
 && A_{y}(x,y) = \sqrt{\frac{4}{5}} \, g \, \cosh(m_V y)  \, \, \eta ^{(1)}(x)+\sqrt{\frac{8}{5}} \, g \, \cosh(m_V y) \sinh(m_V y)  \, \, \eta ^{(2)}(x),
\label{A_sol}  
\eea  
where the gauge coupling in the 4D effective theory is defined as $g=\sqrt{\frac{15 m_V}{16 M}}$. 

\section{Domain-Wall Higgs Field and Higgs Mechanism }
\label{sec:3}
Next we briefly review the 5D extension of the Higgs mechanism 
  which is proposed in Refs.~\cite{DWSM, DWSM2}.  
The reader who is familiar with this subject may skip this section. 
To simplify our discussion, we take the Abelian Higgs model as an example, 
  corresponding to the previous section on the localized U(1) gauge field.  
It is straightforward to extend our discussion to the SM Higgs doublet case. 

For the non-compact 5D spacetime, we need to consider a localization mechanism 
   for not only Higgs field but also its VEV. 
For this purpose, we may utilize the same procedure taken for the gauge field. 
Let us introduce the Lagrangian for the Higgs sector as follows: 
\bea
\mathcal{L}_5^H= g^2 \sy \left[ ({\cal D}^{M}H)^{\dagger}({\cal D}_{M}H)
  -\frac{1}{2}\lambda \left(H^{\dagger}H-\frac{v^2}{2} \right)^2 \right], 
\label{L_H}  
\eea
where $H$ is the Higgs field, $\lambda$ is a Higgs quartic coupling,
  $v$ is its VEV, 
  and the covariant derivative is given by ${\cal D}_M= \partial_M - i Q_H A_M$ 
  with a U(1) charge $Q_H$ for the Higgs field.  
Here, we have taken $s(y)$ to be the same as Eq.~(\ref{exp2}) with $\gamma=3$. 
We have introduced the overall factor $g^2$, by which the zero-mode of the Higgs field 
   is canonically normalized in the effective 4D theory.

Expanding $H$ about the vacuum $H=(v + h + i \phi)/\sqrt{2}$ and neglecting the interaction terms, 
  we obtain (up to total derivative terms)
\bea
\mathcal{L}_5^H & \supset & \frac{1}{2} g^2\sy \left[ (\partial^{M}h)(\partial_{M}h)-m_{h}^{2}h^2\right]
 +\frac{1}{2}g^2\sy(\partial^{M}\phi)(\partial_{M}\phi) \nonumber \\
&=&-\frac{1}{2} g^2s h(\Box_{4}+m_{h}^{2})h+\frac{1}{2}g^2h\,\partial_y(s \partial_{y}h) 
  -\frac{1}{2} g^2s \phi\Box_{4}\phi+\frac{1}{2}g^2\phi\,\partial_{y}(s \partial_{y}\phi) , 
\eea
where $m_h^2= \lambda v^2$ is the physical Higgs boson mass. 
After applying the KK-mode decomposition for these fields, 
\bea
h(x,y)=\sum_{n=0}^{\infty}h^{(n)}(x)\chi^{(n)}_{h}(y), 
  \quad 
\phi(x,y)=\sum_{n=0}^{\infty}\phi^{(n)}(x)\chi^{(n)}_{\phi}(y), 
\label{hfunc}
\eea
   we can see that the KK-mode equations for $\chi^{(n)}_{h}$ and $\chi^{(n)}_{\phi}$ 
   are identical to that of the gauge boson in Eq.~(\ref{KK_EOM}). 
The identical configuration of $\phi^{(0)}$ and $A_\mu^{(0)}$ is from the theoretical consistency
   since the zero-mode $\phi^{(0)}$ is the would-be NG mode eaten by $A_\mu^{(0)}$. 
The KK-mode decomposition of the physical Higgs field is given by (see Eq.~(\ref{A_sol}))
\bea
  h(x,y)=  h^{(0)}(x)  + 2 \sinh(m_V y) \, h^{(1)}(x)\, + \frac{1}{\sqrt{5}} \, (5-4\cosh ^2(m_V y))  \,h^{(2)}(x), 
\label{H_KK}
\eea
with which the free Lagrangian for the scalar fields in the effective 4D theory is given by 
\bea
\mathcal{L}_4^H &\supset& 
 - \frac{1}{2} 
 h^{(0)}\left( \Box_{4} + m_h^2 \right) h^{(0)} 
- \frac{1}{2} 
  \sum_{n=1}^{2} 
\left[ h^{(n)}
  \left( \Box_{4}+ \left(m_h^2 + m_{n}^2 \right) \right) h^{(n)} \right] \nonumber \\
 &&+ \left\{h^{(m)} \to \phi^{(m)}, \; m_h \to 0\right\}.
\eea 
Here, the kinetic terms are canonically normalized.

Associated with the U(1) gauge symmetry breaking by $\langle H \rangle =v/\sqrt{2}$, 
   the U(1) gauge boson acquires its mass. 
After normalizing the kinetic terms for all zero-modes and KK-modes, 
    we find the gauge boson masses \cite{DWSM, DWSM2},
\bea 
   m_A^{(0)} = Q_H  g v,  \quad m_A^{(n)} = \sqrt{m_n^2 + (Q_H g v)^2}, 
\eea
 for the zero-mode and the KK-modes, respectively. 
Note that the formula for the zero-mode gauge boson mass is exactly the same 
  as the one in the 4D  Abelian Higgs model.

The extension of the 5D Abelian Higgs model to the SM case is straightforward. 
For the 5D SM gauge bosons corresponding to the SU(3)$_c \times$SU(2)$_L \times$U(1)$_Y$ gauge groups, 
  we employ the same function $s(y)$ up to a normalization factor which determines the size 
  of each gauge coupling. 
Hence, the KK-mode spectrum for the SM gauge bosons are the same. 
After the electroweak symmetry breaking  we find the gauge boson (photon ($\gamma$), $W$ boson and $Z$ boson) 
  mass spectrum as follows: 
For the zero-modes, 
\bea
   m_\gamma =0, \; \; m_W = \frac{1}{2} g_2 v, \; \; m_Z = \frac{1}{2} g_Z v,   
\eea  
where $g_Z=\sqrt{g_2^2 + g_Y^2}$ with $g_2$ and $g_Y$ being the SU(2)$_L$ and U(1)$_Y$ gauge couplings, respectively, 
   and $v=246$ GeV is the Higgs doublet VEV, 
   while their KK-mode mass spectrum is given by   
\bea
   m_\gamma^{(n)} = m_n, \; \; m_W^{(n)} = \sqrt{m_n^2 + m_W^2}, \; \; m_Z^{(n)} = \sqrt{m_n^2 + m_Z^2}. 
\eea   
The gluon mass spectrum is the same as the photon mass spectrum. 
Since the KK-mode functions for all the SM gauge bosons are identical (up to normalization factors), 
    there is no mass mixing between the KK-modes of photon and $Z$ boson.

\section{Fermion Mass Hierarchy from Geometry }
\label{sec:4}
Finally, we consider localized  fermions in the 5D bulk, whose zero-modes 
   are identified with the SM chiral fermions. 
Following the mechanism proposed in Ref.~\cite{DWF}, 
   we first introduce a real scalar field ($\varphi(x, y)$) in the 5D bulk: 
\begin{eqnarray} 
 {\cal L}_{(5)} = 
  \frac{1}{2} \left(\partial_{M} \varphi \right) \left( \partial^{M} \varphi \right) - V(\varphi)   \; ,  
\end{eqnarray}
where the scalar potential is give by 
\bea 
 V(\varphi) = \frac{m_\varphi^4}{2 \lambda} 
	   -m_\varphi^2 \varphi^2  
           + \frac{\lambda}{2} \varphi^4 .  
\eea
As a solution of the equation of motion, we consider a non-trivial background configuration $\varphi_{\rm kink} (y)$, 
   which is known as the kink solution,
\bea
\varphi_{\rm kink} (y) = \frac{m_\varphi}{\sqrt{\lambda}} 
\tanh [m_\varphi y] .  
\label{kink} 
\eea
Here, we have chosen the kink center at $y=0$.
Expanding the scalar field around the kink background, 
   $\varphi (x, y) = \varphi_{\rm kink}(y) + \tilde{\varphi}(x, y)$, 
   and using the KK decomposition of $\tilde{\varphi}(x, y) = \sum_{n=0}^{\infty} \varphi^{(n)}(x) \chi^{(n)}_{\varphi}(y)$, 
   we obtain the KK-mode equation for $\chi^{(n)}_{\varphi}(y)$. 
It is easy to see that the equation is identical with the first equation in Eq.~(\ref{KK_EOM}),  
   but $s(y)= \left[ \cosh(m_\varphi y) \right]^{-4}$.  
Hence, the canonically normalized KK-mode expansion is found to be  
\bea
   \varphi (x, y) = \varphi_{\rm kink}(y) +  \frac{\sqrt{3 m_\varphi} }{2} \left[ \frac{1}{\cosh^2(m_\varphi y)} \right] \varphi^{(0)}(x) 
    + \sqrt{ \frac{3m_\varphi}{2} } \left[ \frac{\sinh (m_\varphi y)}{\cosh^2 (m_\varphi y)} \right] \varphi^{(1)}(x), 
\label{s-mode}    
\eea  
where $\varphi^{(0)}(x)$ is the massless NG mode corresponding to the spontaneous breaking 
  of the translational invariance in the 5th dimension, 
  and $\varphi^{(1)}(x)$ is the 1st KK-mode with a mass $m_\varphi^{(1)} = \sqrt{3} m_\varphi$.  

We now introduce the Lagrangian for a bulk fermion coupling with $\varphi$, 
\bea
\mathcal{L}&=&i \overline{\psi}\left[\gamma^{\mu}D_{\mu}+i\gamma^{5}D_{y}\right]\psi+Y \varphi \overline{\psi}\psi \nonumber \\ 
&=& 
i \overline{\psi_L} \gamma^{\mu}D_{\mu} \psi_{L}+i \overline{\psi_R} \gamma^{\mu}D_{\mu} \psi_{R}  \nonumber \\
&-& \overline{\psi_L} D_{y}\psi_{R}+\bar{\psi}_{R}D_{y} \psi_L + Y \varphi \left( \overline{\psi_L}\psi_{R}+ \overline{\psi_R}\psi_{L}\right), 
\label{DM_fermion}
\eea
where we have decomposed the Dirac fermion $\psi$ into its chiral components, $\psi= \psi_L + \psi_R$, 
  the covariant derivative is given by $D_M=\partial_M - i g_f A_M^aT^a$ with a SU(N) gauge coupling $g_f$ 
  for $\psi$ in the fundamental representation, 
  and $Y$ is a positive constant.   
Neglecting the gauge interactions and replacing $\varphi$ by the kink background, 
  the equations of motion are given by 
\bea
&& i\gamma^{\mu} \partial_{\mu}\psi_{L}- \partial_{y}\psi_{R} + Y \varphi_0 \psi_{R}=0,  \nonumber \\
&& i\gamma^{\mu}  \partial_{\mu}\psi_{R}+ \partial_{y}\psi_{L} + Y \varphi_0 \psi_{L}=0.
\label{Ferm_EOM}
\eea
Using the KK-mode decompositions, 
\bea
 \psi_L (x,y) = \sum_{n=0}^\infty \psi_L^{(n)}(x) \, \chi_L^{(n)}(y), \; \; \;
 \psi_R (x,y) =\sum_{n=0}^\infty \psi_R^{(n)}(x) \, \chi_R^{(n)}(y), 
\eea
  we have the KK-mode equations as 
\bea
& \left[ -\partial_y^2 - (Y \varphi_{\rm kink})^\prime  +  (Y \varphi_{\rm kink})^2 \right]  \chi_L^{(n)} = m_n^2   \chi_L^{(n)}, \nonumber\\
& \left[ -\partial_y^2 + (Y \varphi_{\rm kink})^\prime  +  (Y \varphi_{\rm kink})^2  \right] \chi_R^{(n)} = m_n^2  \chi_R^{(n)}. 
\label{KKF_EOM}
\eea
These equations are equivalent to the two equations in Eq.~(\ref{KK_EOM4}) 
   by the replacements, 
   $m_\varphi \to m_V$, $Y/\sqrt{\lambda} \to \gamma$, and $\chi_L^{(n)}, \, \chi_R^{(n)} \to \tilde{\chi}^{(n)}, \, \tilde{\psi}^{(n)}$.
Hence, the mass eigenvalues and eigenfunctions are given by Eqs.~(\ref{mn1})-(\ref{sol1}).   
Note that a zero mode only exists for a left-handed fermion in the 4D effective theory. 
This zero-mode is identified with a left-handed SM fermion. 
If instead we flip the sign of $Y$ from positive to negative in Eq.~(\ref{Ferm_EOM}), 
  the right-handed fermion has a zero mode, which is identified with a right-handed SM fermion.
  
For our phenomenology discussion in Sec.~6, let us consider fermions located in close proximity to the kink center.
In this case, the normalized zero mode wavefunction (for a left-handed SM fermion) is approximately described as
\bea
 \psi_L^{(0)}(x,y) = \psi_L(x)\chi_L^{(0)}(y)\approx 
  \psi_L(x)\left[ \left(\frac{m_F^2}{\pi} \right)^{\frac{1}{4}} e^{-\frac{1}{2}m_{F}^2(y-y_{0})^2}\right],
\label{F_example} 
\eea
where $m_F=\sqrt{Ym_{\varphi}^2/\sqrt{\lambda}}$ is the inverse of the DW fermion width,
  and we have set the kink center at $y=y_0$. 
  As discussed in the original proposal of the split fermion scenario of Ref.~\cite{AS}, we introduce a different bulk mass term $M_i$, for each 5D fermion. 
  The setup for all localized SM fermions utilizes the same kink solution, and as a result the kink center is simply shifted by $y_0^i=\frac{M_i}{m_F^2}$.  

\begin{figure}[t]
\begin{center}
\includegraphics[scale=0.86]{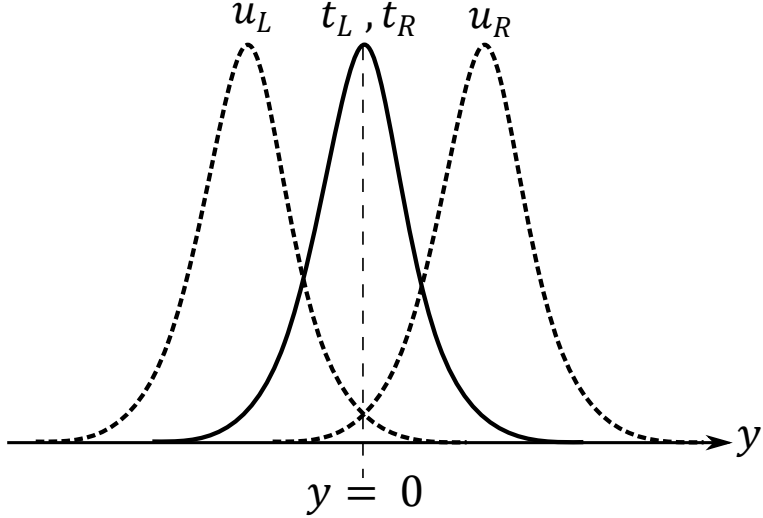} \;
\end{center}
\caption{
The up quark components $u_L$ and $u_R$ are separated by some distance along the 5th dimension, 
   while $t_L$ and $t_R$ are localized at the same location. 
A small overlap between the chiral fermions results in an exponentially suppressed effective Yukawa coupling.   
}
\label{fig:1}
\end{figure}

Let us now extend our system to the SM case. 
We introduce the Yukawa coupling of the SM quarks in 5D as 
\bea
  {\mathcal L}_Y =- Y_u^{ij} \overline{Q}^{i} \tilde{H }u^{j} - Y_d^{ij} \overline{Q}^{i} H d^{j}+{\rm H.c.} 
    \supset - Y_u^{ij} \overline{Q}_L^{i} \tilde{H} u_R^{j} - Y_d^{ij} \overline{Q}_L^{i} H d_R^{j}+{\rm H.c.} , 
\label{Yukawa}  
\eea
where we have decomposed the fields into their chiral components $Q^i=Q_L^i+Q_R^i$, $u^i=u_L^i+u_R^i$, $d^i=d_L^i+d_R^i$ ($i=1,2,3$ is the generation index), 
  and  $H$ ($\tilde{H}=i \sigma_2 H$) is the 5D Higgs doublet. 
With the kink background, zero-modes of $Q_L^i$, and $u_R^i$/$d_R^i$ are identified 
  with left-handed quark doublet and right-handed quark singlets. 
The zero mode wavefunctions for $Q^i$ and $u^i$/$d_R^i$ are given by Eq.~(\ref{F_example}), 
  with the localization positions being different for each SM fermion.

Since the zero mode for the Higgs field and its VEV are flat along the extra dimension 
  ($\chi^{(0)}_{h}(y)={\rm constant}$ in Eq.~(\ref{hfunc})), 
  the effective 4D Yukawa couplings are obtained by integrating out the zero mode quark wavefunctions 
  with respect to the 5th dimensional coordinate. 
For example, the effective Yukawa coupling of the up-type quarks is given by
\bea
Y_{\rm eff}^{ij}= Y^{ij}_u\int_{-\infty}^{\infty}dy  \, \chi^{i (0)}_{L}(y-y_L) \, \chi^{j (0) }_{R}(y-y_{R}),
\label{yeff}
\eea
where the zero modes of the left-handed and right-handed up-type quarks are represented 
    by $\chi^{i(0)}_{L}(y-y_L)$ and $\chi^{j(0)}_{R}(y-y_{R})$, respectively, are localized
    at $y_L$ and $y_R$  in the 5th dimension. 
Note that the effective 4D Yukawa coupling is determined by the amount of overlapping
   between the left-handed and right-handed zero mode wavefunctions. 
Figure \ref{fig:1} illustrates an example between the chiral components of the up and top quarks. 
For the  chiral components of the zero mode wavefunctions of the top quarks localized at $y=0$, they completely overlap with each other, 
   and hence the effective Yukawa coupling is identical to the original Yukawa coupling.    
On the other hand, the chiral components for the zero-mode wavefunctions of the up quarks 
   are localized at different positions, resulting in a highly suppressed effective Yukawa coupling. 
For the present example,  we obtain the effective 4D Lagrangian for the mass terms from Eq.~(\ref{Yukawa}):
\bea
  {\mathcal L}_Y^4  &\supset& -\frac{v}{\sqrt{2}}\left(Y^{33}_u\overline{t_L} t_R+Y^{11}_ue^{- \frac{m_F^2}{4} (\Delta L_{11}^u)^2}\, \overline{u_L} u_R \right. \\ \nonumber 
    &&+ \left. Y^{31}_ue^{- \frac{m_F^2}{4}(\Delta L_{31}^u)^2}\,\overline{t_L} u_R 
  +Y^{13}_ue^{- \frac{m_F^2}{4}(\Delta L_{13}^u)^2}\, \overline{u_L} t_R \right),
\label{Yukawa_KK_Higgs}  
\eea
where $\Delta L_{ij}^u=|y_L^i-y_R^j| $. 
Note that even if $Y_u^{ij}={\cal O}(1)$, the fermion mass hierarchy 
  can be easily reproduced thanks to the geometric factor, $e^{- \frac{m_F^2}{4} (\Delta L_{ij}^u)^2}$.
The origin of the mass hierarchy is now to be interpreted as just a factor difference among the $\Delta L$'s. 

The mass matrices for the up-type and down-type quarks are given by 
   $M_u^{ij}=Y_u^{ij} ( v/\sqrt{2}) e^{- \frac{m_F^2}{4} (\Delta L_{ij}^u)^2}$ and 
   $M_d^{ij}=Y_d^{ij} ( v/\sqrt{2}) e^{- \frac{m_F^2}{4} (\Delta L_{ij}^d)^2}$, respectively. 
The unitary rotations, $u_L^{i} \to U_L^{ij}u_L^{j}$, $d_L^{i} \to V_L^{ij}d_L^{j}$, 
   and similarly for $L \leftrightarrow R$, diagonalize these mass matrices as follows:
\bea
{\rm diag}(m_u,m_c,m_t)&=& U_L^{\dagger}M_uU_R, \nonumber \\
{\rm diag}(m_d,m_s,m_b)&=& V_L^{\dagger}M_dV_R.
\label{mass}
\eea
The quark mixing matrix ($V_{CKM}$) is defined as $V_{CKM}=U_L^\dagger V_L$, and in general $M_u$ and $M_d$ are $3 
\times 3$ complex matrices. 
To simplify our analysis we take both matrices to be Hermitian, so that $U_{L} = U_{R}$ and $V_{L}= V_{R}$, respectively.
We have found numerically that $M_u$ is approximately	 a diagonal matrix, and as a result the rotation matrices are $U_{L} = U_{R} \approx \mathbf{1}$ and $V_{L}= V_{R}\approx  V_{CKM}$.

We fix the top quark Yukawa coupling to be $Y^{33}_u=Y_{t}=0.995$ corresponding to $m_t = 173$ GeV, 
    while we take a mild hierarchy choice and  assign all remaining Yukawa elements as $0.1 \, Y_t$.
The separation distance $\Delta L_{ij}=|y_L^i-y_R^j| $ between the left- and right-handed fermions
    are fixed by their measured masses and found from using Eq.~(\ref{mass}). 
The localization positions of the fields are a free parameter so long as the separation distance is satisfied. 
In our analysis, we have made the choice of left- and right-handed localization positions 
  to be $y_L=\Delta L/2$ and $y_R=-\Delta L/2$. 
For example, the left-handed and right-handed top quarks are localized at $y=0$.
We employ the same procedure for finding the fermion masses of the lepton sector, 
    where the rotation matrices for the charged leptons and neutrinos are, respectively,  
    $\tilde{U}_{L} = \tilde{U}_{R} \approx \mathbf{1}$ and $\tilde{V}_{L}=\tilde{V}_{R}\approx V_{PMNS}$. 
We assume the normal hierarchy of the light neutrino mass spectrum.

\section{Effective Gauge Couplings} 
\label{sec:5}

\begin{figure}[t]
\begin{center}
\includegraphics[scale=0.81]{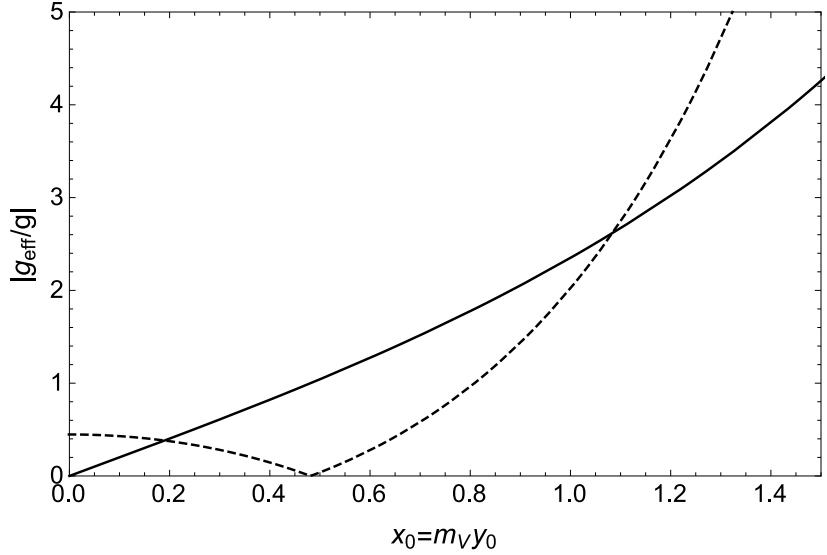} \;
\end{center}
\caption{
Magnitude of effective gauge coupling for the 1st KK (solid line) and the 2nd KK (dashed line) modes 
  as a function of fermion localization position. 
Here, we have taken $\epsilon=1$.
}
\label{fig:2}
\end{figure}

Let us now describe the interaction Lagrangian in the 4D effective theory as 
\bea
 {\mathcal L}_4 \supset  
  \overline{\psi}_L^{(0)} i \gamma^\mu \left( \partial_\mu - i Q_f g A_\mu^{(0)} \right) \psi_L^{(0)} 
+ \sum_{n=1}^\infty  Q_f \, g_{\rm eff}^{(n)} \, A_\mu^{(n)}  \left[\overline{\psi}_L^{(0)} \gamma^\mu \psi_L^{(0)} \right], 
\eea
where the 4D effective gauge coupling between the chiral fermion and the $n$-th KK-mode gauge boson is found 
   by integrating out the zero-mode function of the fermion and the $n$-th KK-mode gauge boson function. 
As an example, the effective gauge coupling of a left-handed SM fermion is given by  
\bea
   g_{\rm eff}^{(n)i}/g =  \int_{-\infty}^{\infty} dy  \left( \chi_{L}^{(0)i}(y-y_L) \right)^2  \chi^{(n)}(y) .
\eea
The effective couplings of the 1st and 2nd KK-modes are given by
\bea 
 {\rm 1st~KK~mode}: \; g_{\rm eff}^{(1)}/g &=& \frac{2}{\sqrt{\pi}} \, m_F 
     \int_{-\infty}^{\infty} dy  \,  \sinh(m_V y)\,e^{-m_F^2 (y-y_L)^{2}}\nonumber \\
   &=&   \frac{2}{\sqrt{\pi}} \,
     \int_{-\infty}^{\infty} dx  \,  \sinh\left(\epsilon x\right)\,e^{-(x-x_L)^{2}} \approx 2 \epsilon x_L ,
     \nonumber \\
{\rm 2nd~KK~mode}: \; g_{\rm eff}^{(2)}/g &=& \frac{1}{\sqrt{\pi}} \, \frac{m_F}{\sqrt{5}} 
     \int_{-\infty}^{\infty} dy  \,  (5-4\cosh(m_V y)^2)\,e^{-m_F^2 (y-y_L)^{2}}\nonumber \\
   &=&  \frac{1}{\sqrt{5\pi}} \, 
     \int_{-\infty}^{\infty} dx  \,  \left(5-4\cosh\left(\epsilon x\right)^2\right)\,e^{-(x-x_L)^{2}} \nonumber \\
   &\approx& \frac{1}{\sqrt{5}}\left(1-2\left(1+2x_L^2\right)\epsilon^2\right) , 
\label{g_eff_formula}     
\eea
where $\epsilon=m_V/m_F $, and we have taken $\epsilon \ll 1$ for the final expression 
  of the two equations in order to simplify the formulas. 
We show these effective couplings in Fig.~\ref{fig:2} as a function of localization position in the bulk, 
  where we have taken $\epsilon=1$ for illustrative purposes. 
For fermions localized at $x_{L}=m_{F}y_{L}=m_{V}y_{L}$ near the origin, 
  the effective gauge coupling for the 1st KK-mode is vanishingly small, 
   since the 1st KK-mode function is an odd function of $y$.
Near the origin, the effective coupling of the 2nd KK-mode attains a non-zero value 
   of $g_{\rm eff}^{(2)}/g \approx 1/\sqrt{5}$, 
   since the 2nd KK-mode function is an even function of $y$.

\section{FCNC Constraints  and KK-mode Phenomenology}
\label{sec:6}
Prediction of the KK-modes in the 4D effective theory is a common property 
  of extra-dimensional models, and we can consider some fascinating phenomenology for them. 
In the Domain-Wall SM, the KK-mode spectra and the coupling manner of each KK-mode 
  with the SM particles depend on the localization mechanism. 
This property is in sharp contrast to, for example, the Universal Extra-Dimension model \cite{UED}, 
  where the KK-mode gauge couplings are universal.  
The localization of the gauge fields offers more variety of the KK-mode phenomenologies 
  than usual compactified extra-dimensional models, 
  thanks to the rich ``geometry" structure of the localized SM particles and their KK-modes. 
In this section, we consider the FCNC constraints on our model and address two scenarios 
  for studying interesting KK-mode phenomenologies: 
  (1) the KK-mode of the SM gauge bosons are extremely heavy and unlikely to be produced at the LHC, 
   while the improved FCNC measurements in the future experiments can reveal the existence 
   of these heavy modes (Sec.~\ref{sec:6.1}). 
  (2) the width of the localized SM fermions is very narrow and as a result, the 4D KK-mode gauge couplings are almost universal. 
  In this case, the FCNC constraints can be easily avoided even for a KK gauge boson mass of order TeV. 
  Such a light KK gauge boson can be searched at the LHC in the near future (Sec.~\ref{sec:6.2}).

\subsection{FCNC Constraints} 
\label{sec:6.1}
In many extra dimensional models (see, for example Ref.~\cite{MK}), 
   a successful localization of the SM fermions to reproduce the fermion mass heirarchy 
   can generate dangerous FCNC processes. 
In the split fermion scenario, Lillie and  Hewett in Ref.~\cite{LH} have considered 
   the FCNC effects mediated by the KK-mode gauge bosons 
   for rare meson decays, and neutral-meson mixings to obtain the constraints 
   on the KK gauge boson masses and fermion localization positions. 
Since the methodology employed in Ref.~\cite{LH} is general, 
   we follow Ref.~\cite{LH} to identify the allowed parameter region for our Domain-Wall SM. 
In our analysis, we employ the updated experimental constraints on FCNCs in Ref.~\cite{PDGFCNC:2018}.

Neglecting the mass splitting induced by the electroweak symmetry breaking (see the last paragraph in Sec.~\ref{sec:3}), 
  the KK-mode spectrum for all SM gauge bosons are the same. 
Thus, we will focus on the KK gluon mediated processes because of its large QCD coupling. 
For the SM left-handed and right-handed quarks, the interactions with the $n$-th KK-mode gluon are given by 
\bea
{\cal L}_{\rm S}&=&
i\bar{u}_{L}^{i}(U_{L}^{\dagger}\tilde{g}^{(n)}_L U_{L})_{ij}\gamma^{\mu}G_{\mu}^{(n)}u_{L}^{j}+
i\bar{d}_{L}^{i}(V_{L}^{\dagger}g^{(n)}_L V_{L})_{ij}\gamma^{\mu}G_{\mu}^{(n)}d_{L}^{j} + (L \leftrightarrow R) \\ \nonumber
&\approx& i\bar{u}_{L}^{i}(\tilde{g}^{(n)}_L)_{ij}\gamma^{\mu}G_{\mu}^{(n)}u_{L}^{j}+
i\bar{d}_{L}^{i}(V_{CKM}^{\dagger}g^{(n)}_L V_{CKM})_{ij}\gamma^{\mu}G_{\mu}^{(n)}d_{L}^{j} + (L \leftrightarrow R),
\eea 
where we have used  $U_{L/R} \approx \mathbf{1}$ and $V_{L/R}\approx V_{CKM}$ in the second line. 
The non-universal KK gluon gauge couplings $g^{(n)}_L$ and $\tilde{g}^{(n)}_L$ are then given by
\bea\label{coupling}
&&\tilde{g}^{(n)}_L={\rm diag}\left\{ g^{(n)}_{eff}(y_{u_L}),g^{(n)}_{eff}(y_{c_L}),g^{(n)}_{eff}(y_{t_L})\right\} \nonumber \\
&&g^{(n)}_L={\rm diag}\left\{g^{(n)}_{eff}(y_{d_L)},g^{(n)}_{eff}(y_{s_L}),g^{(n)}_{eff}(y_{b_L}) \right\}, 
\eea
where the effective couplings are dependent on the localization positions of the left-handed fermions. 
The expressions for $g^{(n)}_R$ and $\tilde{g}^{(n)}_R$ are analogous to Eq.~(\ref{coupling}).
As we have discussed in Sec.~\ref{sec:4}, the localization positions are determined (relatively to a $m_F$ value) 
  so as to reproduce experimental value of the quark masses and their mixing matrix elements. 
Hence, the non-universal KK gluon gauge couplings are determined once $\epsilon=m_V/m_F$ is fixed.

The most stringent bounds on FCNCs come from the meson oscillation measurements found in Ref.~\cite{PDGFCNC:2018}. 
In Ref.~\cite{LH} the authors consider the effective 4-Fermi interactions mediated by the KK gluons, 
  and derive a lower mass bound on the 1st KK gluon ($m_{1}$) as 
\bea\label{mbound}
m_{1} \geq \beta \sqrt{\,F(g_L,g_R)},
\eea
where $\beta$ is a parameter with a mass-dimension of one. $\beta$ is determined by the meson parameters, 
   such as the meson decay constant, the corresponding meson mass, the strong coupling constant, 
   and the meson mass difference.
For example, for the $K^0-\bar{K}^0$ oscillation (which turns out to provide the most stringent FCNC constraint), 
  we find $\beta[{\rm TeV}]=$ $1038$, 
  and $F(g_L,g_R)$ is explicitly given by \cite{LH}
\bea
&&F(g_L,g_R)=\left\{\frac{4}{5} \left|\left(V_{CKM}^{\dagger}\frac{g^{(1)}_L}{g} V_{CKM} \right)_{12} \right|^2
+\frac{1}{40} \left| \left(V_{CKM}^{\dagger}\frac{g^{(2)}_L}{g} V_{CKM} \right)_{12} \right|^2 + (L \leftrightarrow R) \right\}
\nonumber \\
&&+\left(\frac{1}{4}+\frac{3}{4}\frac{m_K^2}{m_d^2+m_s^2} \right)\left\lbrace\frac{4}{5} \left| \left(V_{CKM}^{\dagger}\frac{g^{(1)}_L}{g} V_{CKM} \right)_{21}\times \left(V_{CKM}^{\dagger}\frac{g^{(1)}_R}{g} V_{CKM} \right)_{12} \right|
\right. \nonumber \\
&&\left. +\frac{1}{40} \left| \left (V_{CKM}^{\dagger}\frac{g^{(2)}_L}{g} V_{CKM} \right)_{21}\times 
\left(V_{CKM}^{\dagger}\frac{g^{(2)}_R}{g} V_{CKM} \right)_{12} \right| 
+ (L \leftrightarrow R) \right\rbrace,
\label{Feq}
\eea
where the subscripts indicate the matrix elements. 
Looking back at the effective couplings in Eq.~(\ref{g_eff_formula}), 
    one can easily understand that the terms $\propto |g^{(1)}_{L/R}|^2$ are sub-leading 
    compared to the terms $\propto |g^{(2)}_{L/R}|^2$ for $\epsilon \ll 1$. 
From numerical calculations, we find that the second term on the right-hand side dominates.

\begin{figure}[t]
\begin{center}
\includegraphics[scale=0.9]{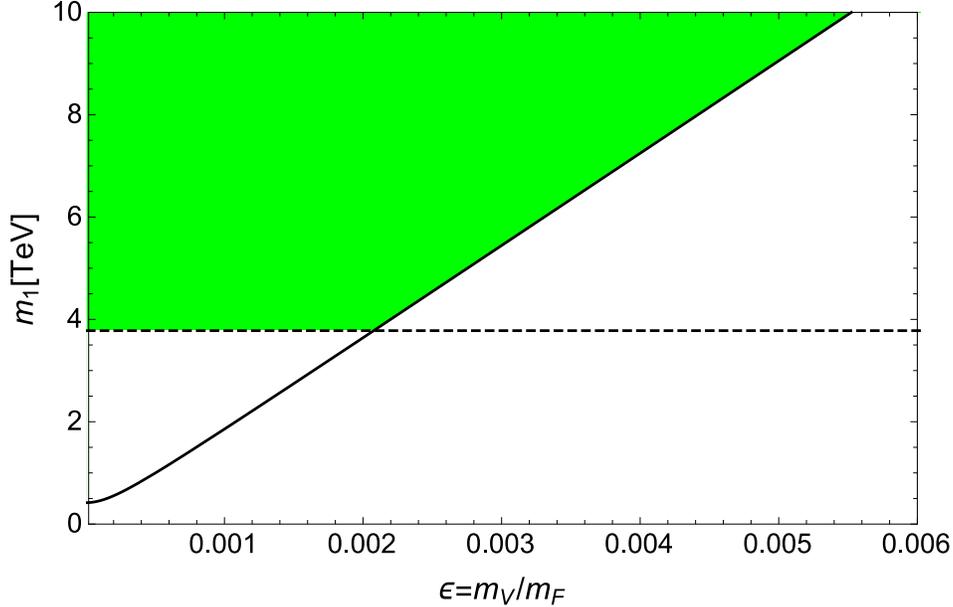}
\end{center}
\caption{
The lower bounds on the 1st KK gluon mass ($m_1$) from the meson oscillation data (diagonal solid line) 
   and the LHC Run-2 results (horizontal dashed lines) as a function of $\epsilon=m_V/m_F$. 
Combining the two constraints, the allowed region is identified as the green shaded region. 
}
\label{fig:3}
\end{figure}

Figure \ref{fig:3} shows the lower mass bounds on the 1st KK gluon as a function of $\epsilon$.  
The diagonal solid line is the constraint obtained from the $K^0-\bar{K}^0$ oscillation. 
We have found that the bounds from the $B^0_d$ and $B^0_s$ measurements are significantly weaker 
   than the $K^0-\bar{K}^0$ constraint and we do not show them in the figure. 
Although our model is not the same as the one examined in Ref.~\cite{LH}, 
   our result is qualitatively consistent with the result presented in Ref.~\cite{LH}. 
For $\epsilon \lesssim 0.002$, Fig.~\ref{fig:3} shows that the FCNC constraints can be avoided even for $m_1 ={\cal O}$(TeV). 
On the other hand, for a larger $\epsilon$ value, the KK gluon must be extremely heavy 
   and far beyond the LHC search reach. 
The horizontal dashed line is the current lower bound from the LHC Run-2  results. 
See the next section for details on this constraint.  

Typically, rare meson decays place tight constraints on extra dimension models as well, 
  however, in our scenario we have found that these effects are an order of magnitude weaker 
  than the $K^0-\bar{K}^0$ oscillation constraint shown in Fig.~\ref{fig:3}. 
The largest such contribution comes from the process $Br(B^0_s \to \mu^+ \mu^-) = 3 \times 10^{-9} $ \cite{LHCb:2017}, 
  from which we have obtained a mass bound on $m_{1} \gtrsim 400$ GeV for $\epsilon=0.004$.

Another possible avenue for detecting flavor violating effects is through the rigorously explored neutron electric dipole moment (EDM). 
Current experiments have put a bound on the neutron EDM at $|d_n/e| <0.3\times 10^{-25}$ cm \cite{EDMexp}. 
 It is well known that SM quantum corrections to the neutron EDM occur only at three loop order, and results in a contribution much smaller than what is currently experimentally accessible at a value of around $|d_n/e| \leq 10^{-30}$  cm \cite{EDMSM1,EDMSM2}.
 Beyond SM physics on the other hand, may introduce larger contributions to the neutron EDM. In particular, flavor violating interactions can culminate in significant deviations from current experimental bounds \cite{EDMBSM}.
 Following the analysis by the authors from Ref.~\cite{EDMth}, we find that our prediction for the EDM when $\epsilon \ll 1$ is 
\bea
|d_n/e| \sim \frac{\alpha_s}{160\pi}\left(\frac{m_d}{m_V^2}\right)\simeq 2.2\times 10^{-26} {\rm cm} \left(\frac{{\rm TeV}}{m_V}\right)^2,
\eea
where $\alpha_s=g_s^2/4\pi=0.118$ is the strong coupling constant, and $m_d=4.7$ MeV is the down quark mass. 
This relation places a lower bound on the 1st KK gluon mass of $m_1 =\sqrt{5}m_V=1.9$ TeV, which is weaker than the LHC bound discussed in the next section. 

\subsection{LHC Phenomenology} 

\begin{figure}[h]
\begin{center}
\includegraphics[scale=1.4]{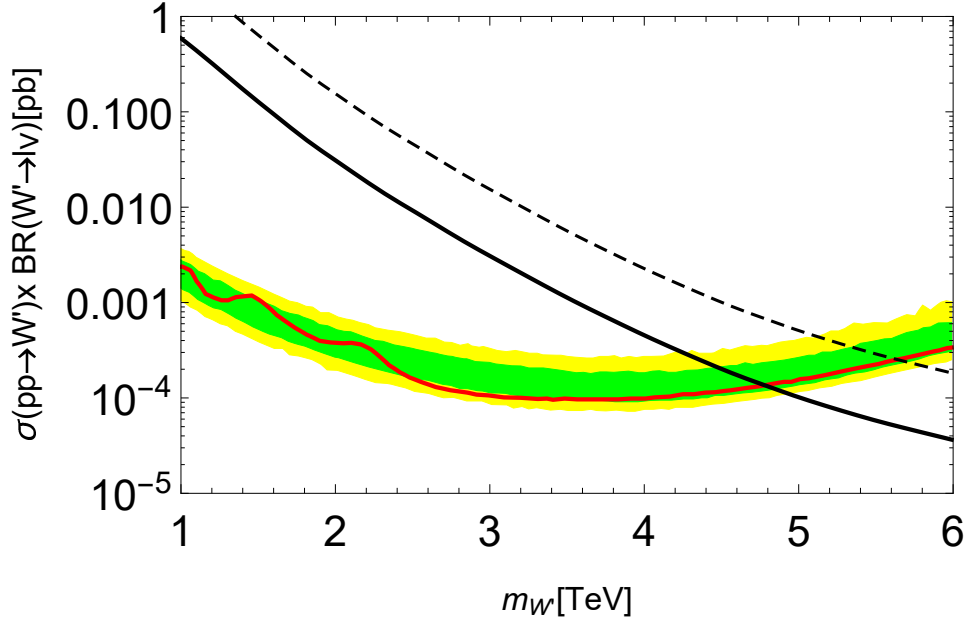}       
\end{center}
\caption{
The cross section times its branching ratio, $\sigma (pp \to W^{(2)} \to l \nu)=\sigma(pp \to W^(2)) {\rm BR}(W^{(2)} \to l \nu))$, 
  as a function of $m_{W^\prime}=m_2$ for $g_{\rm eff}^{(2)}/g=1/\sqrt{5}$ (solid diagonal line), 
  along with the theoretical prediction of $\sigma (pp \to W^\prime \to l \nu)$ for the sequential SM $W^\prime$ boson (dashed diagonal line) 
  and the cross section upper bound at 95\% Confidence Level  (solid horizontal curve in red) 
  from the analysis by the ATLAS collaboration with the integrated luminosity of 79.8 ${\rm fb}\,^{-1}$ \cite{ATLAS2:2019}. 
Here, the $1 \sigma$ (green) and $2 \sigma$ (yellow) expected limit bands are also shown. 
}
\label{fig:4}
\end{figure}

\label{sec:6.2}
The ATLAS and the CMS collaborations have been searching for new particle resonances 
  with a variety of final states at the LHC Run-2.  
For the sequential SM, $Z^\prime$ and $W^\prime$ bosons have the same properties as the SM $Z$ and $W$ bosons except for their masses. 
The ATLAS (CMS) collaboration has recently reported their search results with luminosity of about 139 ${\rm fb}^{-1}$ (36 ${\rm fb}^{-1}$) for the $Z^\prime$ search, and 80 ${\rm fb}^{-1}$ (36 ${\rm fb}^{-1}$) for the $W^\prime$ search.
The ATLAS (CMS) lower bound on the sequential SM $Z^\prime$ boson is obtained to be $m_{Z^\prime} \geq 5.1\,(4.5)$ TeV 
  with dilepton final states \cite{ATLAS1:2019} (\cite{CMS1:2018}),   
  while $m_{W^\prime} \geq 5.6\,(5.2)$ TeV for the sequential SM $W^\prime$ boson mass is obtained 
  with its decay mode $W^\prime \to l \nu$  \cite{ATLAS2:2019} (\cite{CMS2:2019}).
  
 Since we have set a common mass $m_V$ for the $y$-dependent SM gauge couplings, 
  the KK gauge boson mass spectra for gluon, photon, weak bosons are approximately the same 
  for $m_{W,Z}^2 \ll m_1^2=5m_V^2$.   
Thus, we consider the most severe constraint from the $W^\prime$ boson search. 
Since the total decay width of $W^\prime$ boson is about 3\% of its mass for $m_{W^\prime} \gtrsim 1$ TeV, 
   we employ the narrow-width approximation in evaluating the parton-level cross section of the process, 
\bea 
   \hat{\sigma}(q \overline{q}^\prime \to W^\prime) \propto \Gamma_{W^\prime}(W^\prime \to q \overline{q}^\prime)  
      \, \delta(M_{\rm inv}^2 -m_{W^\prime}^2)  \propto  g^2,  
\eea   
where $M_{\rm inv}^2$ is the invariant mass of the initial partons, 
   $\Gamma_{W^\prime}(W^\prime \to q \overline{q}^\prime)$ is the partial decay width into $q \overline{q}^\prime$, 
   and $g$ is the SM SU(2) gauge coupling. 
When we identify the $W^\prime$ boson with the $n$-th KK-mode of the SM $W$ boson in the Domain-Wall SM,  
   the only difference is from the effective gauge coupling $g_{\rm eff}^{(n)}$. 
We have found from Fig.~\ref{fig:3} that $\epsilon \ll1$ is required to avoid the FCNC constrains for $m_1={\cal O}$ (TeV).  
Since $g_{\rm eff}^{(1)} \propto \epsilon$ for $\epsilon \ll 1$, 
   the 1st KK-mode $W$-boson production cross section is very small. 
On the other hand, the 2nd KK-mode $W$-boson has a sizable coupling, $g_{\rm eff}^{(2)}/g \approx 1/\sqrt{5}$,
   for $\epsilon \ll 1$, and we use the LHC Run-2 result to obtain a lower mass bound on the 2nd KK mode. 
Now, we have a relation, 
\bea
   \sigma (pp \to W^{(2)} \to l \nu) = \left( \frac{g_{\rm eff}^{(2)}}{g} \right)^2 \sigma (pp \to W^\prime \to l \nu) 
     \simeq  0.2 \times \sigma (pp \to W^\prime \to l \nu), 
\eea
where $\sigma (pp \to W^\prime \to l \nu)$ is the cross section for the sequential SM $W^\prime$ boson.

In Fig.~\ref{fig:4}, we show the cross section $\sigma (pp \to W^{(2)} \to l \nu)$ as a function of $m_{W^\prime}=m_2$
  for the value of $g_{\rm eff}^{(2)}/g=1/\sqrt{5}$ (solid diagonal line), 
  along with the upper bound on the cross section from the ATLAS results \cite{ATLAS2:2019} 
  at the LHC Run-2 with a 79.8 fb$^{-1}$ integrated luminosity (horizontal solid curve in red)
  and the theoretical prediction of $\sigma (pp \to W^\prime \to l \nu)$ for the sequential SM $W^\prime$ boson (dashed line). 
We can read off the lower bound on the 2nd KK-mode mass ($m_2$) 
  from the intersection of the corresponding solid diagonal line and the solid horizontal (red) curve, 
  which is $m_2[{\rm TeV}]\geq 4.78$. 
Considering the ratio between the 1st and 2nd KK gauge boson masses as $m_1/m_2=\sqrt{5/8}$, 
   we obtain the lower bound on the 1st KK gauge boson mass to be $m_1[{\rm TeV}]\geq 3.78$. 
This result is shown in Fig.~\ref{fig:3} as the horizontal dashed line.

Combining the constraints from the FCNC processes and the LHC Run-2 results, 
   we find the allowed parameter region shown as the green shaded region in Fig.~3. 
This figure shows two typical regions: 
  (1) $\epsilon \gtrsim 0.002$, for which the FCNC constraints are more severe then the LHC constraint. 
  (2) The LHC constraint is more severe for $\epsilon \lesssim 0.002$. 
Interestingly, the FCNC and the LHC constraints are complementary to constrain 
  the model parameter space of the Domain-Wall SM.
  
 Another phenomenological signal worth considering is the process $W^\prime \to jj$, which enables us to place a mass constraint on the KK gluons. 
 In Ref.~\cite{ATLAS3:2017} the ATLAS collaboration searched for the $W^\prime$ resonance signal and placed a lower mass constraint of $3.6$ TeV (seen in their Fig.~3(c)). 
 This process can be used to place a constraint on the KK gluon mass in our model by using the relation $(\frac{g_s}{g})^2 \times \sigma \times A \times BR \approx (4) \times \sigma \times A \times BR$, where $g_s$ and $g$ are the strong and weak gauge couplings respectively and A is the scaling parameter that is typically less than one. From this relationship we find that the lower bound on the KK gluon mass is around $m_2[{\rm TeV}] \sim 4.2$. As we have shown above, this is a weaker constraint than the one we found for the $W^\prime$ boson.

\section{Summary and Discussions}
Recently in Refs.~\cite{DWSM, DWSM2}, 
  we have proposed a framework of the 5D Domain-Wall SM 
  and investigated its phenomenology, where all the SM fields are localized in certain 3-dimensional domains 
  in non-compact 5D spacetime. 
In this paper, we have extended our previous work to naturally explain 
  the mass hierarchy among the SM fermions by using the idea of the split fermion scenario. 
Although the fermion mass hierarchy problem can be solved with a mild hierarchy
  among the model parameters, our model is subject to very severe FCNC constraints. 
In addition, the current LHC Run-2 narrow resonance search results
  provide a lower mass bound on the KK-mode SM gauge bosons. 
Considering the FCNC and the LHC constraints, we have arrived 
  at two typical cases for phenomenological viability of our model: 
   (1) the KK-mode of the SM gauge bosons are extremely heavy and unlikely to be produced at the LHC, 
      while future FCNC measurements can reveal the existence of these heavy modes. 
   (2) the width of the localized SM fermions is very narrow, leading to almost universal 4D KK-mode gauge couplings.  
   In this case, the FCNC constraints can be easily avoided even if a KK gauge boson mass lies at the TeV scale. 
   Such a light KK gauge boson can be searched at the LHC Run-3 and then the High-Luminosity LHC 
   in the near future.
Interestingly, the current experimental constraints from the FCNC measurements 
   and the LHC search for the KK-mode gauge bosons are complementary 
   to constrain parameter space of the Domain-Wall SM.

Since gravitons reside in the bulk, we also need to consider a localization of graviton 
  to complete our proposal of the Domain-Wall SM. 
For this purpose, we may combine our scenario with the RS-2 scenario \cite{RS2} 
  with the Planck brane at $y=0$. 
Here we may identify the Planck brane as a domain-wall with the zero-width limit. 
The mass spectrum of the KK-modes of the SM fields is controlled 
  by the width of the domain-walls, and the current LHC results constrain it to be $\lesssim$(1 TeV)$^{-1}$. 
On the other hand, the width of 4-dimensional graviton is controlled by the AdS curvature $\kappa$ 
   in the RS-2 scenario and its experimental constraint is quite weak, $\kappa \gtrsim 10^{-3}$ eV \cite{RS2}. 
Therefore, we can take $\kappa \ll 1$ TeV and neglect the warped background geometry 
  in our setup of the Domain-Wall SM, while the 4-dimensional Einstein gravity is reproduced in the RS-2 scenario 
  at low energies. 
We may think if the energy density from the SM domain-walls is large and affects the RS-2 background geometry. 
However, we expect the energy density from the domain-walls of ${\cal O}(\Lambda^4)$ with $\Lambda={\cal O}$(1 TeV), 
  while the energy density of the Planck brane in the RS-2 scenario is given by ${\cal O}(M_P^2 \kappa^2)$ 
  with the reduced Planck mass of $M_P\simeq 2.4 \times10^{18}$ GeV.  
Therefore, we choose the AdS curvature in the range of $10^{-3}$ eV$\ll \kappa \ll$1 TeV for the theoretical consistency of our scenario.
 
\section*{Acknowledgements}
This work is supported in part by the United States Department of Energy Grant (DE-SC0012447
 (N.O. and D.V.) and DE-SC0013880 (D.R.)).

{}

\end{document}